\documentclass[12pt]{iopart}
\usepackage{graphicx}

\newcommand{\beq}{\begin{equation}}
\newcommand{\eeq}{\end{equation}}
\newcommand{\bea}{\begin{eqnarray}}
\newcommand{\eea}{\end{eqnarray}}
\newcommand{\ba}{\begin{array}}
\newcommand{\ea}{\end{array}}
\newcommand{\bc}{\begin{center}}
\newcommand{\ec}{\end{center}}

\newcommand{\half}{\hbox{$1\over2$}}
\newcommand{\fourth}{\hbox{$1\over4$}}
\newcommand{\eight}{\hbox{$1\over8$}}
\newcommand{\commentout}[1]{{}}

\begin{document}

\title{Measurement theory for spinor condensates}

\author{Juha Javanainen}

\address{Department of Physics, University of Connecticut, Storrs, CT
06269-3046, USA}

\begin{abstract}
We study the experimental signatures of several states of a Bose-Einstein
condensate of spin-1 atoms by quantum trajectory simulations of
Stern-Gerlach experiments. The measurement process itself creates an
apparent random alignment for the spins, so that it is difficult to
distinguish between a condensate that initially has an alignment in an
unknown direction and one with no alignment at all. Nonetheless, in
repeated experiments with identically prepared condensates, it is possible
to discriminate between certain ground states of the condensate proposed in
the literature.\end{abstract}

\pacs{03.75.Fi,03.65.Bz}


\maketitle
\section{Introduction}
With the introduction of the optical trap in the experiments on
Bose-Einstein condensation, simultaneous confinement of atoms in several
angular momentum states is now feasible~\cite{STA}. The atomic spins
interact in spin-exchange  collisions, which gives rise to collective
many-spin states. Energy differences between several conceivable ground
states of the spinor condensate~\cite{HO98,LAW98,HO00} have turned out to
be small, so that it need not be obvious which one gets prepared in an
experiment.

On the other hand, resolving the contradicting opinions of the
past~\cite{AND86,LEG91}, recent analyses have shown~\cite{QPH,QTS1,QTS2}
that a Bose-Einstein condensate may display signatures of spontaneously
broken gauge symmetry, such as interference of two independently prepared
condensates~\cite{KET97} or the Josephson effect, even if the gauge
symmetry of the state is in fact not broken. The lesson is
that the relation between the state of the system and the measurement
results may be conditioned as much by the experiment as by the state
itself. How to distinguish experimentally between the proposed states of
the spinor condensate~\cite{LAW98,HO00} is therefore a nontrivial
question in its own right.

In this paper we study the experimental signatures of a spin-1 condensate
for three different initial states, ``coherent state'' with spontaneous
alignment~\cite{HO00}, ``fragmented state'' in which there are two
macroscopically occupied spin states~\cite{HO00}, and rotationally
invariant ``singlet state''~\cite{LAW98}. Our main tool is quantum
trajectory simulations~\cite{ORQTS,QTS1,QTS2} of repeated Stern-Gerlach
experiments. Once more~\cite{QPH,QTS1,QTS2}, the interplay between state
and measurement is evident. It appears that the coherent and
fragmented states are experimentally indistinguishable, but can be
distinguished easily from the singlet state. We also present a testable
prediction for the distribution of the outcomes of the Stern-Gerlach
experiments that probes the statistic of spin projections to all orders,
not just averages and standard deviations. Finally, as an important
by-product of our development we note that, while the simulations at
first sight only apply to a quite artificial experimental scheme, they
implement what we believe is a universal theory for spin measurements on
a condensate.

\section{States of spinor condensate}
We consider putative ground states for the spinor condensate in the
case when the spin species do not separate, but the atoms with
the three angular momentum projections $m=\pm1$ and $m=0$ commingle with
identical spatial wave functions. We assume scattering lengths such that
the interaction between the spins is effectively antiferromagnetic. For
brevity we assume zero magnetic field. Nonetheless, we still pick a
quantization axis $z$. We write the annihilation operators for the atoms
with the three spin components along the quantization axis as $a_\pm$ and
$a_0$. For notational simplicity, we take the total number of atoms $N$
to be even.

The coherent state is arrived at via the semiclassical argument,
which treats the annihilation operators $a_\pm$ and $a_0$ as
$c$ numbers~\cite{HO00}. The $c$ numbers come with phases that are not
uniquely determined by minimization of energy.  Assigning a value for
the phases is tantamount to inserting by hand a form of spontaneous
symmetry breaking. The corresponding quantized ansatz for the state
of the spinor condensate reads~\cite{HO00}
\beq
|C\rangle = {1\over\sqrt{2^N\,N!}}\left(e^{-i\chi} a^\dagger_- + e^{i\chi}
a^\dagger_+\right)^N|{\rm vac}\rangle\,.
\eeq
Positing the state $|C\rangle$ is analogous to assuming that a
single-component condensate is in a coherent state, as opposed to, say, a
number state of the atoms~\cite{QPH,QTS1,QTS2}. $\chi$ is an angle that
characterizes the symmetry breaking. It selects preferred directions in
the $xy$ plane. 

The fragmented or ``coherent-fragmented state'' \cite{HO00} is
just a number state with half of the spins in the state $+$ and half in
the state $-$,
\beq
|F\rangle = |N/2,0,N/2\rangle\,.
\eeq
The arguments give the numbers of atoms with
the $z$ components of the angular momentum equal to $-1$, $0$,
and $+1$. In the thermodynamic limit the state $|F\rangle$ should model the
ground state in the presence of even the most minute magnetic field in the
$z$ direction~\cite{HO00}. While this state could behave differently in the
$x$ and $z$ directions, it has no built-in structure in the $xy$ plane.

The singlet state of \cite{LAW98} is of the form
\beq
|S\rangle = \sum_{k=0}^{N/2}A_k|k,N-2k,k\rangle\,.
\eeq
The
coefficients $A_k$ satisfy the recursion relation
\beq
A_k = -\sqrt{N-2k+2\over N-2k +1}\,A_{k-1}\,,
\eeq
and are then fixed by normalization, except for an arbitrary
overall phase. The state $|S\rangle$ is the unique ground state for
exactly zero magnetic field. It has the same form in all rotated
frames, and thus possesses no intrinsic direction.

\section{Measurements on spins}
In our model for the measurements, we  assume for the time being that
one atom at a time is removed from the condensate and is subject to a
Stern-Gerlach experiment that probes the spin component in the $xy$
plane in a fixed direction at an angle $\varphi$ with respect to the $x$
axis. The observed sequence of spin projections is recorded and
analyzed. All measurements are supposed to be completed in a time shorter
than the evolution time scale of the spins in the condensate owing to
spin-spin interactions.

From angular momentum algebra it is easy to see that the annihilation
operators for the spin states in the direction $\varphi$ are
given in terms of the original quantization axis operators as
\bea
a_\pm(\varphi) &=& {e^{i\varphi}\over2}\,a_-\pm{1\over\sqrt{2}}\,a_0+
{e^{-i\varphi}\over2}\,a_+\,,\\
a_0(\varphi) &=& {e^{i\varphi}\over\sqrt{2}}\,a_- -
{e^{-i\varphi}\over\sqrt{2}}\,a_+\,.
\eea
We carry out conventional quantum trajectory simulations of the
measurement sequence~\cite{QTS1,QTS2} numerically. Thus, suppose that,
entering the $n$th measurement,  the normalized state of the spins is
$|\psi_n\rangle$. We first calculate the probabilities for each spin
projection,
\beq
P^{(n)}_m = {\langle\psi_n| a^\dagger_m(\varphi)a_m(\varphi)
|\psi_n\rangle\over N-n}
\label{PROB}
\eeq
for $m=-1$, $0$ and $+1$. Second, we use a random number generator to
pick one of the $m$ values in such a way that the probability for
choosing $m$ equals $P_m^{(n)}$. Third, we reduce the wave packet
according to the selected value of $m$, so that the state vector for the
$(n+1)$th measurement is
\beq
|\psi_{n+1}\rangle = {a_m(\varphi)|\psi_n\rangle\over
\sqrt{\langle\psi_n| a^\dagger_m(\varphi)a_m(\varphi)
|\psi_n\rangle}}\,.
\label{NEWSTATE}
\eeq
The choices of the spin projection $m$ for each measurement $n$ constitute
the data.

We first consider the coherent state, $|\psi_0\rangle=|C\rangle$. It
turns out that, except for an inconsequential overall phase, and
independently of the past choices of the spin projections, the state
vector $|\psi_n\rangle$ is always a coherent state of the same form as
$|C\rangle$, except that of course the total number of atoms is $N-n$ not
$N$. Furthermore, the probabilities for the outcomes of the Stern-Gerlach
experiment are always the same,
\beq
P_\pm^{(n)} = \half\cos^2(\varphi-\chi),\quad P_0^{(n)} =
\sin^2(\varphi-\chi)\,.
\eeq
Successive Stern-Gerlach experiments are therefore uncorrelated. The
nature of the symmetry breaking is alignment. Namely, in the measurement
directions
$\varphi=\chi\pm\half\pi$ one only sees the result $m=0$, whereas in the
directions $\varphi=\chi$ and $\varphi=\chi+\pi$ one finds
$m=\pm1$ with equal probabilities, and no $m=0$.

Now, according to the notion of spontaneous symmetry breaking, the angle
$\chi$ varies at random from one spinor condensate to the next, and the
experimenter has no a priori way of knowing it. As the observed frequencies
of spin projections $N_0$, $N_\pm$ vary wildly with the angle $\chi$,
measurements on just a single condensate do not seem to make a
particularly discriminating test for or against the state
$|C\rangle$.

To gather more incisive data, we imagine repeating the experiment
with a large number of identically prepared condensates. In principle we
should discuss the frequencies of the observed spin projections, but in
our simulations we have access to the probabilities
$P^{(n)}_0$ and
$P^{(n)}_\pm$ as well. For brevity, here we often pick a particular
measurement
$n$, and focus on the combination of the probabilities
\beq
P = P_-^{(n)}+P_+^{(n)}-P_0^{(n)}\,.
\label{PTEST}
\eeq
For a given condensate in the coherent state $|C\rangle$ this has
the value $P=\cos^2(\varphi-\chi)-\sin^2(\varphi-\chi)$, but the
angle $\chi$ is unknown. Nonetheless, if the angle $\chi$ is evenly
distributed over the unit circle $[0,2\pi)$, we may calculate the
 probability density $f(P)$ for the values of $P$. The result is
\beq
f(P) = \left\{\begin{array}{ll}
{\displaystyle1\over\displaystyle\pi\sqrt{1-P^2}},&P\in(-1,1);\\
0,& {\rm otherwise}\,.
\end{array}\right.
\label{DIST}
\eeq

Thus, we envisage using a large number of condensates. For each
condensate one extracts the combination of probabilities $P$, in our
numerical experiments using (\ref{PTEST}) for some given measurement
$n$, and in real experiments by estimating the probabilities using
the observed frequencies of spin projections for each condensate. Finally
one compares the distribution of the numbers $P$ with predictions such as
(\ref{DIST}). In the case of the coherent state
$|C\rangle$, the numerical experiments simply amount to testing of our
algorithms.

\begin{figure}
\begin{center}\includegraphics[scale=1.00]{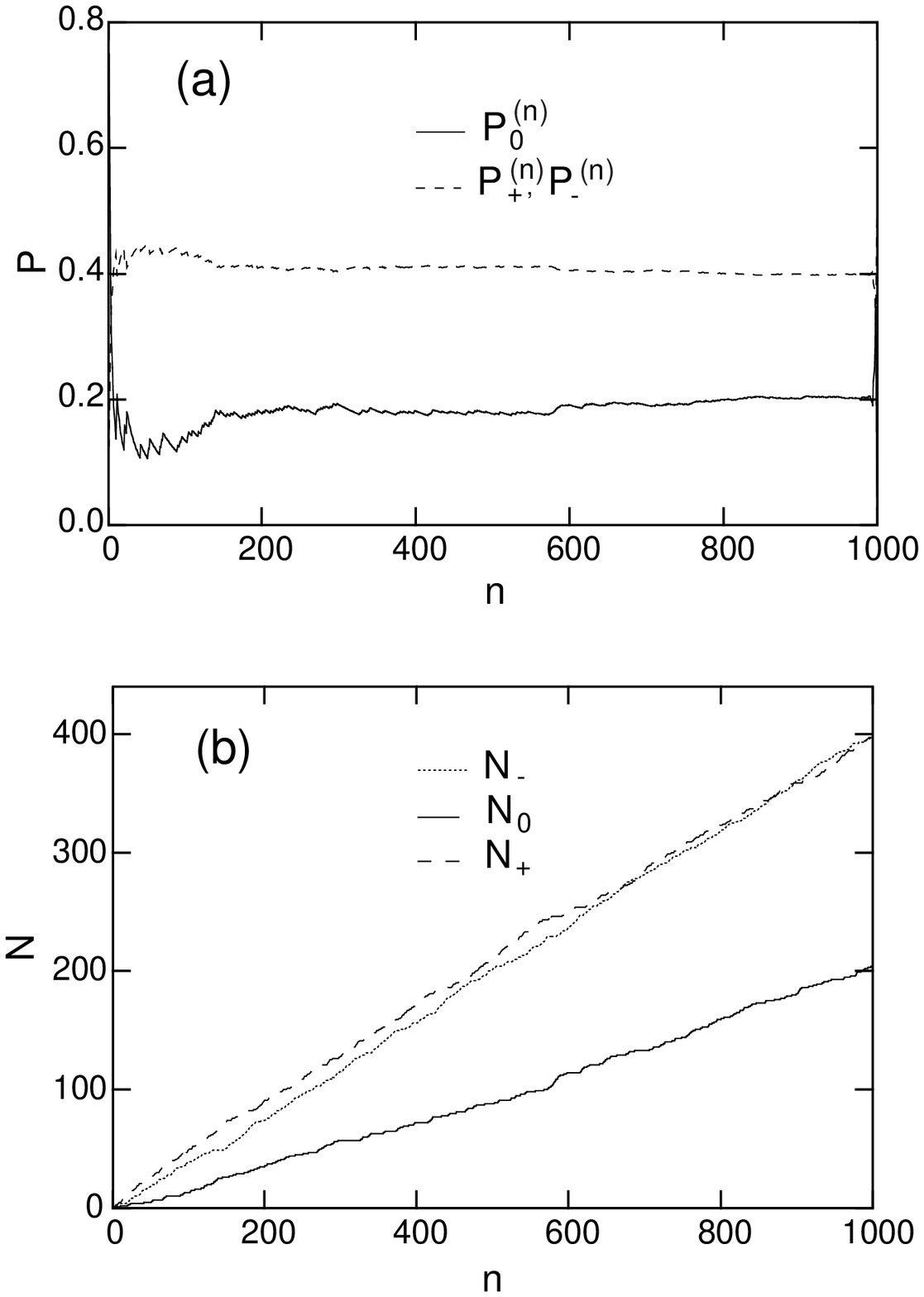}\end{center}
\caption{Simulated probabilities (a) and cumulative frequencies (b) for
spin projections as a function of measurement number in one particular
condensate that starts out with 1000 atoms in the fragmented state
$|F\rangle$.}
\label{ONERUN}
\end{figure}

We next move on to the fragmented state $|F\rangle$.  In
figure~\ref{ONERUN}(a) we plot the numerically simulated probabilities
$P_0^{(n)}$ and
$P_\pm^{(n)}$ as a function of the measurement number $n$ for a sample
condensate with
$N=1000$ atoms, and figure~\ref{ONERUN}(b) gives the corresponding
cumulative frequencies $N_0$, $N_\pm$. Notably, after an initial
transient, the probabilities approximately stabilize at values compatible
with the observed relative frequencies of the spin components.

The expectation values for the numbers of spins with given projections
are $\langle a^\dagger_\pm(\varphi) a_\pm(\varphi)\rangle=\fourth N$,
$\langle a^\dagger_0(\varphi) a_0(\varphi)\rangle =\half N$. One
might naively expect that, even in a single condensate, one would find
the spins approximately in the ratios $N_-:N_0:N_+=1:2:1$. This clearly
is not the case. In figure~\ref{ONERUN}(b) the ratios rather are 2:1:2.
Moreover, the observed ratios vary at random from one condensate to
another.

The measurements are correlated; the outcome of an observation
of the projection of a spin affects the state, which in turns affects the
outcome of future observations. The correlations work out in an
interesting way. In the first measurement the probabilities are
$\fourth:\half:\fourth$, but then they quickly drift away and stabilize
at some other values. Once the probabilities have stabilized,
Stern-Gerlach measurements seem to be approximately independent
repetitions of one another. However, the apparent state that is being
repeatedly measured varies at random from one condensate to the other.
Measurements on the fragmented state $|F\rangle$ in an individual
condensate behave remarkably like measurements on the symmetry-broken
coherent state $|C\rangle$.

\begin{figure}
\begin{center}\includegraphics[scale=1.00]{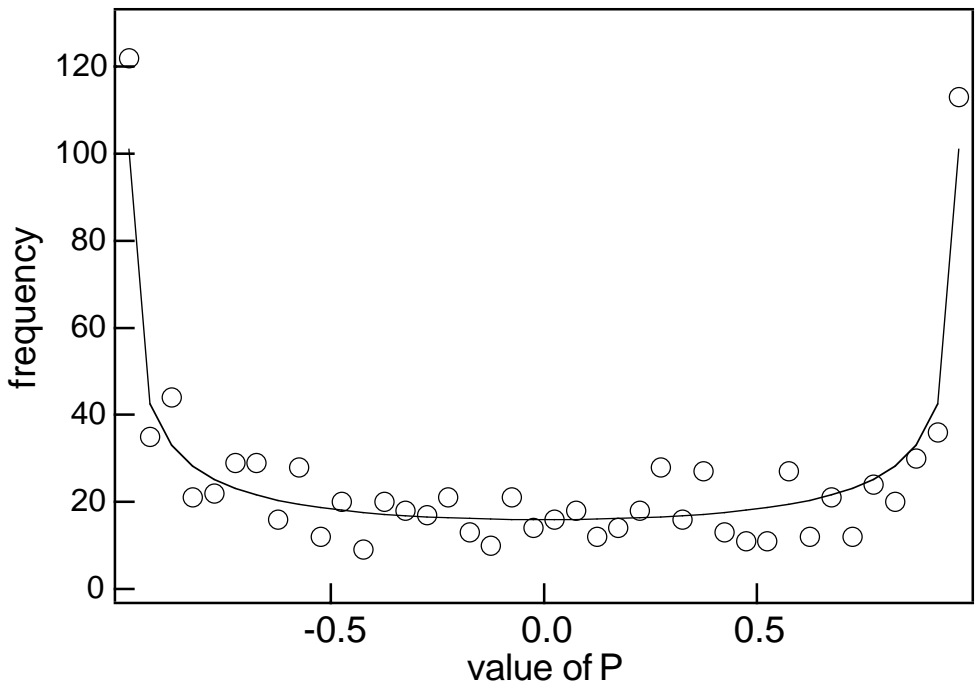}\end{center}
\caption{Histogram of the values $P$ as in (\protect\ref{PTEST})
from numerical simulations of the fragmented state
$|F\rangle$ (circles), and as deduced from (\protect\ref{DIST})
for the coherent state $|C\rangle$ (solid line). 1000
independently prepared condensates with
$N=500$ atoms each are used in the simulations. The range $(-1,1)$ of the
quantity
$P$ is divided into 40 equally wide bins, and the vertical axis shows the
number of $P$ values in each bin. We use the measurement $n=101$ in
the simulation data.}
\label{HISTOGRAM}
\end{figure}

It is legitimate to ask if the states $|F\rangle$ and
$|C\rangle$ can be distinguished at all. We address this question
further in figure~\ref{HISTOGRAM}. We take 1000 spinor condensates, each
with
$N=500$ atoms. For each condensate we record the quantity $P$ of
(\ref{PTEST}) for the Stern-Gerlach measurement number $n=101$. We
then bin the results into 40 equally wide slots of $P$, and draw the
histogram as circles. Also shown as the solid line is the prediction for
the histogram if the probability distribution for the values of $P$ is
given by (\ref{DIST}), as appropriate for the coherent state
$|C\rangle$. Given the statistical fluctuations, we cannot tell the states
$|F\rangle$ and
$|C\rangle$ from one another.

The situation is remarkably similar to the one that was encountered
earlier with a single-component condensate when it came to the question
of a number state, versus a coherent state whose phase arises from
spontaneously broken gauge symmetry~\cite{QPH,QTS1,QTS2}. When detection of
the atoms is explicitly considered, it turns out that both states
lead to interference between independently prepared condensates, even
though the number state seemingly does not provide any phase for the
interference~\cite{QPH,QTS1}. Both coherent-state and number-state
condensates even exhibit the Josephson effect~\cite{QTS2,JAV90}. It is
possible in principle to tell the difference between a coherent state and
a number state by studying atom statistics for a small number (at most
tens) of atoms~\cite{JAV97}, but an experiment discriminating between
coherent and number states is yet to be carried out in a condensate of any
size. We believe that, analogously, there would be little or no practical
difference between the states
$|C\rangle$ and
$|F\rangle$ of the spinor condensate.

\begin{figure}
\begin{center}\includegraphics[scale=1.00]{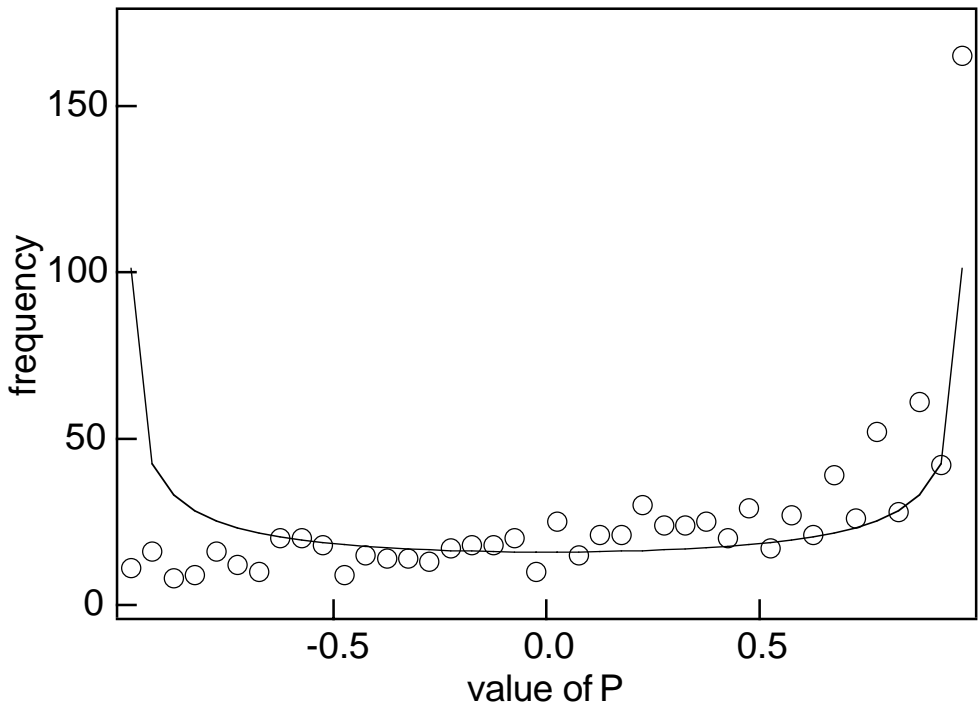}\end{center}
\caption{Same as figure~\protect\ref{HISTOGRAM}, except that the singlet
state $|S\rangle$ was used in the simulations.}
\label{HISTOGRAMS}
\end{figure}

Let us finally consider the singlet state $|S\rangle$. The expectation
values of the spin components are all $\langle a^\dagger_\pm(\varphi)
a_\pm(\varphi)\rangle=\langle a^\dagger_0(\varphi)
a_0(\varphi)\rangle={1\over3}N$. However, just like in the case of the
fragmented state $|F\rangle$, in a single condensate both the simulated
probabilities and the frequencies stabilize at values that may be
completely different from $N_-:N_0:N_+$ = $1:1:1$. One
may, of course, draw a histogram such as in figure~\ref{HISTOGRAM} also
for the state
$|S\rangle$, which we have done in figure~\ref{HISTOGRAMS}. The
discrepancy with the prediction of the coherent state $|C\rangle$ is
evident even without any statistical analysis.

Drawing the histogram for the state $|S\rangle$ might already have been be
a case of overkill. While the frequencies for spin projections are
seemingly random for a single condensate, we have found that they behave
as expected when averaged over many condensates:  $1:2:1$ for the states
$|C\rangle$ and
$|F\rangle$, $1:1:1$ for the state $|S\rangle$. The singlet $|S\rangle$
should be experimentally distinguishable from the other two states simply
by averaging the frequencies of spin projections over many condensates.

\section{Comparing schemes for Stern-Gerlach experiments}
Up to now we have envisaged picking one atom at a time out of the
condensate. In reality one might rather do an en masse
Stern-Gerlach experiment,  subjecting the entire spinor condensate at
once to a gradient of the magnetic field. The spin projections then
separate in space, and can be counted by counting atoms.

We assume that the
detection is in principle still one atom at a time, as if the atoms fell on
small detectors like photons hitting silver bromide grains on a
photographic film. It is known from the theory of photon
detectors~\cite{ATS} that the joint counting rate of an array of
absorbing broadband detectors is proportional to normally ordered 
correlation functions of electric field operators. Analogously, we
conjecture that the joint probability of seeing a sequence of spin
projections
$m_1, m_2,\ldots,m_N$ is given by
\beq
P(m_1,\ldots,m_N) = K \langle
a^\dagger_{m_1}(\varphi)\ldots a^\dagger_{m_N}(\varphi)
a_{m_N}(\varphi)\ldots a_{m_1}(\varphi)
\rangle\,,
\label{PROBALL}
\eeq
where $K$ is the normalization constant needed to make the sum of all
probabilities equal to unity. In photon detection theory the electric
field operators are in time order as well, but inasmuch as we assume that
we may ignore the reversible evolution of the spins during the
measurement, the annihilation (creation) operators in (\ref{PROBALL})
are taken at the same time and commute. The joint probabilities are then
independent of the order of the indices $m_1,\ldots,m_N$, i.e.,
independent of the order in which the atoms are detected.

Suppose that the spin system starts out with the state vector
$|\psi_0\rangle$, and consider the correlation functions as in
(\ref{PROBALL}). We may, of course, write
\bea
&&\langle\psi_0| a^\dagger_{m_1}(\varphi)\ldots a^\dagger_{m_N}(\varphi)
{a_{m_N}(\varphi)\ldots a_{m_1}(\varphi)|\psi_0 \rangle
}\nonumber\\
&&=\left[\langle\psi_0|
a^\dagger_{m_1}(\varphi)\right]\,a^\dagger_{m_2}(\varphi)\ldots
a^\dagger_{m_N}(\varphi)a_{m_N}(\varphi)\ldots a_{m_2}(\varphi)\,
\left[a_{m_1}(\varphi)|\psi_0\rangle\right]\,.
\eea
Trivial as this rewrite is, it immediately leads to two crucial
observations. First, in (\ref{PROBALL}) the probabilities for the
various outcomes $m_1$ are precisely as given by
(\ref{PROB}) for $n=1$. Second, once any particular value $m_1$ has been
picked, the process starts over; we are left with the correlation
functions
\[
\langle\psi_1|\,a^\dagger_{m_2}(\varphi)\ldots
a^\dagger_{m_N}(\varphi)a_{m_N}(\varphi)\ldots a_{m_2}(\varphi)
\,|\psi_1\rangle\,
\]
where the new state $|\psi_1\rangle$ is the $n=1$ version of
(\ref{NEWSTATE}). We have thus shown that our quantum trajectory
simulations produce spin sequences $m_1,\ldots,m_N$ that have the
probabilities predicted by (\ref{PROBALL}). A fortiori, the results from
quantum trajectory simulations agree with the predictions of
(\ref{PROBALL}).

Our simulations were seemingly for a scheme in which one
atom at a time is subject to a Stern-Gerlach experiment. We did not offer
any concrete suggestions on how to carry out such experiments, but
neither did we have to. Inasmuch as our reasoning leading to
(\ref{PROBALL}) is correct, our simulations also model the {\em
usual\/}~\cite{STA} Stern-Gerlach measurements on the spinor condensate.

One might also study the number operators for
different spin projections, such as $n_0(\varphi) =
a^\dagger_0(\varphi)a_0(\varphi)$, in total
disregard of the measurement process. As far as we can tell, our quantum
trajectory simulations averaged over many condensates give the
same qualitative and semi-quantitative results that one would expect on
the basis of elementary quantum mechanics with the operators
$n_0(\varphi)$, $n_\pm(\varphi)$.

As a qualitative example, the
fluctuations of $n_0(\varphi)$ in the state
$|F\rangle$ are large, $\Delta n_0 \simeq\sqrt{\eight}\,N$.
In quantum trajectory simulations one correspondingly sees wildly varying
ratios in the frequencies of spin projections between individual
condensates. From this viewpoint, measurement-induced alignment is the
reason for the large fluctuations in $n_0(\varphi)$ [and
$n_\pm(\varphi)$]. More quantitatively, one may compute the expectation
value and the standard deviation for the number of atoms with a given
spin projection in a condensate either by using quantum mechanics with
the operators
$n_0(\varphi)$, etc., or by averaging over quantum trajectory simulations
of a large number of condensates. We have found that, for large
$N$, the results agree.

\section{Concluding remarks}
Both quantum trajectory simulations and (\ref{PROBALL}) give seemingly
sensible predictions for the frequencies of spin projections and their
fluctuations even for the en masse Stern-Gerlach experiment. But
they do a lot more, too. For instance, a histogram such as in
figure~\ref{HISTOGRAM} may be prepared for the results
of en masse experiments and compared with the predictions of
quantum trajectory simulations. Such a histogram probes the spin
statistics in principle to all orders, beyond average and standard
deviation. We thus have a nontrivial prediction from (\ref{PROBALL}) that
may be contrasted with experiments. Another possible extension of our
method is time evolution, as in the Josephson effect~\cite{QTS2} or in
the time dependence of spin correlations. We may straightforwardly
incorporate time ordering and time evolution of the annihilation
operators into (\ref{PROBALL}). In contrast, an approach in terms of the
number operators $n_0(t)$, $n_\pm(t)$ without an explicit consideration of
the measurements probably makes a subtle affair~\cite{JAV90}.

It would seem worthwhile to derive (\ref{PROBALL}) from microscopic
arguments~\cite{GOL98}, along the lines of earlier work on
photons~\cite{ATS}.  To the extent that this can be done, our quantum
trajectory simulations would be proven to be a universal method for
the studies of the detected properties of a spinor condensate. More
generally, coherence theory of light is built entirely on the theory of
photon detection, so it appears plausible that the eventual coherence
theory for Bose-Einstein condensates is based on detection theory as well.
The combination of measurement theory and accompanying quantum trajectory
simulations is at present our best bet for a general approach to the
coherence properties of a condensate.

In sum, we have laid down a measurement theory and the accompanying
quantum trajectory simulations for a spinor condensate, and proposed a
nontrivial experiment in which our approach may be tested. The main result
thus far is that the state of the spinor condensate and measurements
thereof are intertwined to the extent that it is probably impossible to
distinguish experimentally between the symmetry-broken coherent state and
the fragmented state of the condensate. In the process we have uncovered
new aspects of measurement theory which should prove fruitful in future
discussions of the coherence properties of a condensate.

\ack

This work is supported in part by NSF, Grant No. PHY-9801888, and by NASA,
Grant No.\ NAG8-1428.

\end{document}